\documentclass[10pt,a4paper,twocolumn]{IEEEtran} 
\normalsize
\usepackage[colorlinks,
linkcolor=black,
anchorcolor=black,
citecolor=black,
urlcolor=black,
]{hyperref}
\usepackage[hyphenbreaks]{breakurl}
\usepackage{diagbox}
\usepackage[english]{babel}
\usepackage{latexsym}
\usepackage{varioref}
\usepackage{mathrsfs}
\usepackage{amssymb}
\usepackage{array}
\usepackage{hhline}
\usepackage{dcolumn}
\usepackage{tabularx}
\usepackage{algorithm}
\usepackage{algorithmic}
\usepackage{calc}
\usepackage[usenames]{color}
\usepackage{amsmath,amssymb,bigdelim}
\usepackage{amsfonts}
\usepackage{t1enc}
\usepackage{graphicx}
\usepackage{pstool}
\usepackage{boxedminipage}
\usepackage{indentfirst}
\usepackage{booktabs}
\usepackage[noadjust]{cite}
\usepackage{caption}
\usepackage{multirow}
\usepackage{subfigure}
\usepackage[usenames]{color}
\usepackage{clrscode}
\usepackage{url}
\usepackage{setspace}
\usepackage{xcolor}
\usepackage{fancybox}
\makeatletter
\def\url@leostyle{%
  \@ifundefined{selectfont}{\def\UrlFont{\sf}}{\def\UrlFont{\small\ttfamily}}}
\makeatother
\newcommand{\rem}[1]{}

\newcommand{\addnew}[1]{{\color{black} #1}}

\renewcommand{\baselinestretch}{1}

\providecommand{\texlivekeywords}[1]{\textbf{\textit{Index terms---}}#1}
\begin{document}

	\pagestyle{plain}
  \title{Toward 6G with Terahertz Communications: Understanding the Propagation Channels}
	\author{
     Xuesong Cai,~\IEEEmembership{Senior Member,~IEEE}, Xiang Cheng,~\IEEEmembership{Fellow,~IEEE}, and Fredrik Tufvesson,~\IEEEmembership{Fellow,~IEEE}
		 \thanks{
     The final version can be found in IEEE Communications Magazine. Digital Object Identifier: 10.1109/MCOM.001.2200386
     
    This work has been funded by the Horizon Europe Framework Programme under the Marie Skłodowska-Curie grant agreement No. 101059091, the Swedish Research Council (Grant No. 2022-04691), the strategic research area ELLIIT, Excellence Center at Linköping – Lund in Information Technology, and Ericsson. 


      X. Cai and F. Tufvesson are with the Department of Electrical and Information Technology, Lund University, Lund, Sweden (email: xuesong.cai@eit.lth.se, fredrik.tufvesson@eit.lth.se). 

   X. Cheng is with the State Key Laboratory of Advanced
Optical Communication Systems and Networks, School of Electronics, Peking
University, Beijing, 100871, P. R. China (email: xiangcheng@pku.edu.cn).
     
     }
     \vspace{-0cm}
}

\markboth{IEEE Transactions on Communications}%
{Submitted paper}

\maketitle \thispagestyle{plain}
\begin{abstract}

  This article aims at providing insights for a comprehensive understanding of terahertz (THz) propagation channels. Specifically, we discuss essential THz channel characteristics to be well understood for the success of THz communications. The methodology of establishing realistic and 6G-compliant THz channel models based on measurements is then elaborated on, followed by a discussion on existing THz channel measurements in the literature. Finally, future research directions, challenges and measures to enrich the understanding of THz channels are discussed.

\end{abstract}
\texlivekeywords{6G, terahertz communication, propagation, channel sounding, parameter estimation, clustering, modeling, spatial consistency, and machine-learning.}
\IEEEpeerreviewmaketitle

\section{Introduction}

While fifth-generation communications have become a commercial reality, research on sixth-generation (6G) wireless systems towards 2030 is now in the center of attention. 
The ITU-T envisions that 6G will drive a high-fidelity holographic society, connectivity for all things and time-sensitive applications that will dramatically change the human society. 
Along with new physical layer techniques and higher layer capabilities, even larger system bandwidths are required in 6G to meet the corresponding requirements in data rates, life-critical latency and reliability. 
The new THz band from 0.1 to 10\,THz has been identified as the key enabler for extreme data rates. Compared to frequency bands below 100\,GHz, the D-band (110-170\,GHz) can, e.g., provide much wider bandwidths and is promising for many advanced applications such as wireless backhaul, virtual reality, localization, etc. 
Consequently, THz communications have attracted a great deal of attention and strategic research initiatives around the world \cite{zhang2020channel}. 
\addnew{For example, in September 2017, the European Union launched a cluster of projects to study THz communications.\footnote{\addnew{See https://thorproject.eu/links/ict-09-2017-cluster/}}} In February 2018, the Federal Communications Commission (FCC) launched a new initiative to expand access to spectrum above 95\,GHz, and in April 2019 designated 21.2\,GHz of unlicensed THz spectra distributed between 100\,GHz and 250\,GHz. The 2019 World Wireless Communications Conference (WRC-19) identified the 275-450\,GHz band for terrestrial mobile and fixed services, among others.

As a first step to design any new generation system, it is essential to investigate
the propagation channels because they are distinct at different frequency bands and fundamentally constrain
the system design. The propagation research aims to measure propagation channels (channel sounding), extract channel
parameters such as delays, angles-of-arrival/departure of multipath components (MPCs), etc. from measurement data (parameter estimation), and build mathematical representations of propagation channels (channel
modeling). In the first-generation systems, the attention was
focused on the power domain, i.e., path loss, shadowing and small scale fading. As the wireless systems evolved, various
techniques such as multiple-access, multiple-antenna, advanced coding, and modulation schemes emerged. It was then
necessary to model the propagation channels in a more refined and comprehensive manner, for the development and
realistic performance evaluation of system proposals and applications. This further involves the parameter domains of
delay, angle, Doppler frequency, polarization. As positioning and sensing are expected to be integrated into
6G, the new channel models must further represent the physical environment in more detail and provide spatial consistency. All the requirements and details make it difficult to
establish realistic THz channel models solely relying on simulations, e.g., ray-tracing \cite{8674541}. 
Measurement-based THz work is hence indispensable to understand the THz channels. 


This article focuses on measurement-based investigations towards a comprehensive understanding of THz propagation channels. 
We first discuss THz applications and channel characteristics that are critical for the success of these applications. Then, we elaborate the main procedures of establishing realistic THz channel models based on measurements. Moreover, existing THz channel measurements in the literature are discussed. Finally, we elaborate future research directions, where challenges and ways forward are also mentioned. 

\section{Requirements for THz channel investigations}

For different applications, the corresponding THz systems require different knowledge of THz propagation channels. The evolution of technology also raises new requirements. Therefore, here we first discuss propagation basics in the THz frequency range, applications and scenarios, and techniques that may be applied therein. After that, we discuss essential THz propagation properties that need to be well understood for the success of THz applications.


\subsection{THz frequency range}

\begin{figure}
  \centering
  \psfrag{Spreading, 1 m}[l][l][0.6]{Free-space,\,1\,m}
  \psfrag{Spreading, 10 m}[l][l][0.6]{Free-space,\,10\,m}
  \psfrag{Spreading, 100 m}[l][l][0.6]{Free-space,\,100\,m}
  \psfrag{Absorption, 1 m}[l][l][0.6]{Absorption,\,1\,m}
  \psfrag{Absorption, 10 m}[l][l][0.6]{Absorption,\,10\,m}
  \psfrag{Absorption, 100 m}[l][l][0.6]{Absorption,\,100\,m}
  \psfrag{Frequency [THz]}[c][c][0.8]{Frequency [THz]}
  \psfrag{Loss [dB]}[c][c][0.8]{Loss [dB]}
  \includegraphics[width=0.47\textwidth]{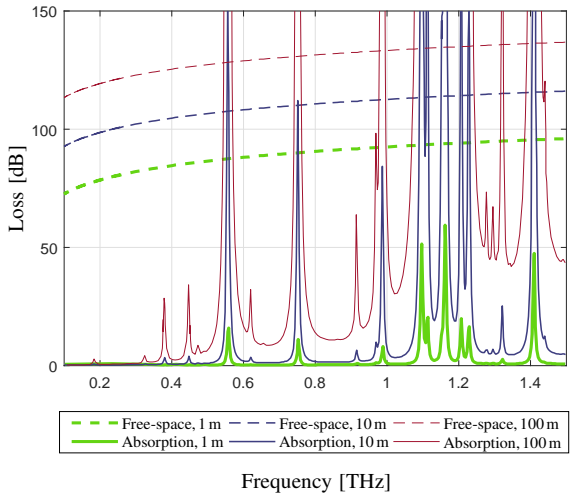}
  \caption{Free-space path losses and molecular absorption losses from 0.1 to 1.5\,THz for different link distances \cite{rajatheva2020scoring}.\label{fig:thzpathloss}}
 \end{figure}

Based on physical assumptions \cite{rajatheva2020scoring}, Fig.\,\ref{fig:thzpathloss} illustrates the obtained free-space path losses and molecular absorption losses in the frequency range from 0.1 to 1.5\,THz. 
The free-space path loss is getting larger with increasing frequency since the effective area of an isotropic antenna decreases. This means that high-gain antennas or large-scale antenna arrays, i.e., ultra-massive multiple-input multiple-output (UM-MIMO) \cite{9216613}, are needed to compensate the severe power loss. 
Moreover, accurate and fast beam-alignment and beam failure recovery are crucial. 
Another critical factor is that the molecular absorption loss is becoming non-negligible at THz frequencies at larger distances and is highly dependent on frequency. It makes thus sense to first exploit the transmission windows centered around, e.g., 150\,GHz, 250\,GHz and 650\,GHz as they offer much lower molecular absorption losses. 

 \begin{table*}[]
  \centering
    \caption{THz applications and required knowledge of channel properties.}
    \scalebox{1}{\footnotesize
    \begin{tabular}{lp{4cm}p{4cm}p{4cm}p{4cm}}
    \hline
   {Category} & Point-to-point communications & Mobile communications & Sensing and localization  \\ \hline
   Specific scenarios &  Intra-device communication, wireless backhaul/fronthaul, etc. & WiFi access, virtual reality, cellular communications, vehicular communications, etc. &  Velocity sensors, THz imaging, simultaneous localization and mapping, etc. \\
                        Techniques &  High gain antennas, UM-MIMO, distributed MIMO, beamforming & High gain antennas, UM-MIMO, distributed MIMO, beam management &  High gain antennas, UM-MIMO, distributed MIMO, beam management \\ 
                        Link distance  &  Short/long-distance  &   Short/long-distance   &  Short/long-distance     \\ 
                        Mobility    &  No  &  Low/mid/high   &  Low/mid/high     \\ 
                        Essential knowledge required  &  Power behavior, weather condition, MPCs, spherical-wavefront effects  & Power behavior, weather condition, MPCs, spherical-wavefront effects, spatial-consistency, dynamic evolution, birth-death behavior  & Power behavior, weather condition, MPCs, spherical-wavefront effects, spatial-consistency, physical scatterers, dynamic evolution, birth-death behavior  \\ \hline 
    \end{tabular}}
    \label{tab:summarization}
    \end{table*}

\subsection{THz applications}

  \textit{Point-to-point (P2P) communications}: The application scenarios include cableless intra-device data transmission, wireless backhaul/fronthaul, data exchange within data centers, etc. They usually require very high throughput on the order of Gbps to avoid cumbersome cable connections. Besides an ultrawide bandwidth, either antennas with very high gains or UM-MIMO are required to achieve high link-SNRs, especially when the link distance is long, e.g., for wireless backhaul/fronthaul. Although there may exist reflection, scattering and blockage, the THz propagation channels are in general quite stable and static with a dominant line-of-sight (LoS) path. However, for outdoor P2P communications, power losses due to different weather conditions can affect the performance significantly since the size of the precipitations are comparable to the THz wavelengths.

\textit{THz mobile communications}: The low-mobility scenarios include WiFi access, cellular communication, proximity communication, on-body communication, data transmission between devices, virtual reality, etc., where communication nodes can move at low to mid velocities. The corresponding THz channels can be challenging, e.g., due to abrupt blockage. Consequently, beam tracking and beam recovery are critical features to enable a stable connection. Another promising technology, the so-called distributed MIMO or cell-free MIMO, can also be implemented. This means that a large number of antennas/antenna-arrays are distributed over a large area. The blockage can be mitigated by redundant antennas. 
\addnew{However, due to the distributed implementation and short symbol duration \cite{9047145}, synchronization, channel estimation, coherent transmission and reception among distributed nodes can be a true challenge.} 
For high-mobility scenarios such as vehicle-to-vehicle and vehicular-to-infrastructure communications, the requirement for beam management is even more challenging due to the highly dynamic THz propagation effects. 

\textit{THz sensing and localization}: To name a few application cases, THz band is suitable for detecting small velocity changes because the Doppler frequency scales with the carrier frequency. Benefitting also from large-bandwidths and UM-MIMO, i.e., high delay and spatial resolutions, the THz band is also promising to realize high-resolution imaging systems \cite{9482537} for, e.g., security checks. The link distance is usually not large, and human body effect is an essential part of the propagation channels. 
Similarly by exploiting the ultra-wide bandwidth and ultra-large antenna arrays, THz systems can efficiently resolve and identify MPCs and thus detect the accurate locations of scatterers/virtual anchors in the environment to achieve highly accurate simultaneous localization and mapping. It is worth noting that the sensed environment information can in turn facilitate THz communications. For example, with a prior knowledge of scatterers' locations, much faster beam recovery can be realized.




\subsection{Essential THz channel properties} 

Like any other wireless systems, link budget calculations of THz systems are fundamental and require modelling of path loss, shadowing and small scale fading. Further, the behavior of MPCs caused by reflection, diffraction and diffuse scattering, and the distributions of MPCs in delay and angular domains need to be investigated.
Whether the THz channels exhibit sparsity (a small number of MPCs) is closely related to whether beamforming or spatial multiplexing or both can be utilized. The exploitation of UM-MIMO and distributed MIMO also necessitates the understanding of spherical-wave propagations since the array aperture can be very large and scatterers can be near to the array. For the same reasons, the channels observed at different parts of the array can be different. This means that the knowledge of the evolution of MPCs across the array is required. In mobile THz communication scenarios, understanding the continuous change of MPCs along time/movement-trajectory, is further needed for the design and evaluation of beam-management techniques. Moreover, the evolution properties of physical scatterers must be reflected in the THz channel models to make them applicable for sensing and localization applications. A brief summary is shown in Table\,\ref{tab:summarization}.


\section{Methodology of Measurement-based THz channel investigations \label{sec:measurement-based}}

Measurements are indispensable to obtain realistic THz channel characteristics. Calibration of simulation-based modeling tools such as ray-tracing also relies on measurements.
Here, we discuss the main procedures of investigating THz channels by measurements, including channel sounding, parameter estimation, clustering and modeling.

\subsection{Channel sounding}

\begin{figure}
  \centering
  \psfrag{f}[c][c][0.6]{$f$}
  \psfrag{t}[c][c][0.6]{$t$}
  \psfrag{Tx}[c][c][0.6]{Tx}
  \psfrag{Rx}[c][c][0.6]{Rx}
  \psfrag{CK}[c][c][0.6]{CLK}
  \textcolor{black}{
      {
  \subfigure[]{\includegraphics[width=0.23\textwidth]{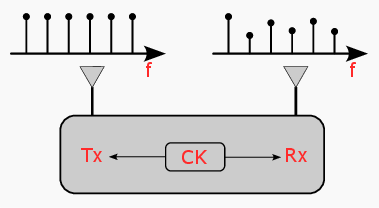}}
  }}
  \psfrag{f}[c][c][0.6]{$f$}
  \psfrag{t}[c][c][0.6]{$t$}
  \psfrag{Tx}[c][c][0.6]{Tx}
  \psfrag{Rx}[c][c][0.6]{Rx}
  \psfrag{CLK}[c][c][0.4]{CLK}
  \subfigure[]{\includegraphics[width=0.23\textwidth]{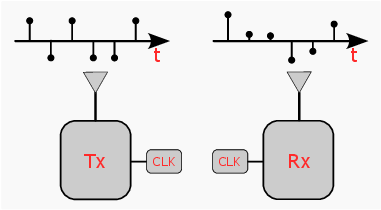}}

  \psfrag{RF}[r][r][0.5]{RF}
  \subfigure[]{\includegraphics[width=0.15\textwidth]{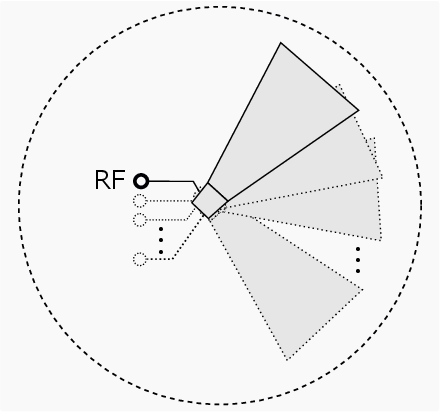}}
  \psfrag{RF1}[c][c][0.5]{$\text{RF}_1$}
  \psfrag{RF2}[c][c][0.5]{$\text{RF}_2$}
  \psfrag{RFN}[c][c][0.5]{$\text{RF}_N$}
  \subfigure[]{\includegraphics[width=0.15\textwidth]{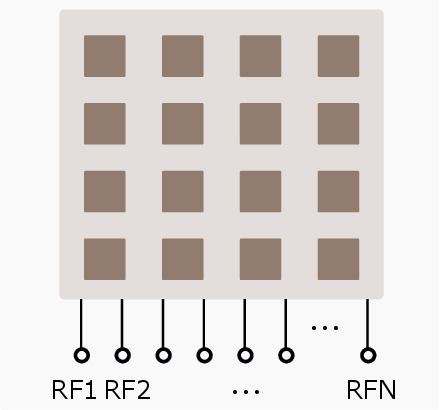}}
  \psfrag{switch}[c][c][0.5]{Switch matrix}
  \psfrag{RFC}[l][l][0.5]{RF}
  \subfigure[]{\includegraphics[width=0.15\textwidth]{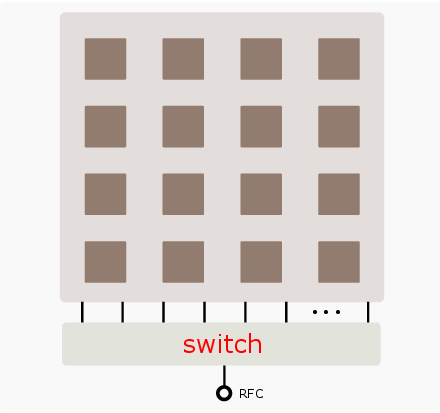}}
  \caption{The main components of THz channel sounders. (a) VNA-based channel sounding. (b) Time-domain channel sounding. (c) A virtual antenna array. (d) A real antenna array. (e) A switched antenna array.\label{fig:thz_sounding}}
 \end{figure}

Meticulously designed measurement systems, i.e., channel sounders, are needed to measure THz propagation channels. 
As illustrated in Fig.\,\ref{fig:thz_sounding}, there are mainly two types of THz channel sounders. 
The first type is based on a vector network analyzer (VNA) and works in the frequency domain \cite{9790802}. 
By sweeping the frequency band of interest using narrowband signals, channel transfer functions are measured. Then the THz channel impulse responses can be acquired by applying inverse discrete Fourier transform to the measured channel transfer functions. 
The advantages of VNA-based channel sounders include that a large THz bandwidth can be easily measured and that the system response of the VNA can be well-calibrated and decoupled. 
However, it usually takes a long time for a frequency-sweep, which limits its use in dynamic scenarios. Moreover, the Tx and Rx have to be connected to the same VNA with cables. The high attenuation of THz signals in the cables limits the distance between Tx and Rx significantly. Therefore, VNA-based THz channel sounders are in general only applicable for short-range static THz channels. 

Other type of THz channel sounders works in the time domain, where real-time waveforms/sequences such as the pseudo-noise sequences and various chirp-like signals occupying a certain bandwidth are transmitted to excite the THz channel \cite{8255203}. 
The channel impulse responses are estimated by performing cross-correlation between the received and transmitted signals. 
The delay resolution of channel impulse responses is determined by the autocorrelation function of the transmitted sequence. 
By modifying the parameters of the sequence, its peak-to-average-power ratio can be controlled for efficient usage of power amplifiers. 
Time-domain channel sounders are more suitable to measure dynamic THz channels. 
Moreover, the Tx and Rx can be synchronized over the air using two pre-synchronized reference clocks, e.g., rubidium clocks, and thus separated for long-distance measurements. 
However, the received waveforms should be sampled at a rate no less than twice the signal bandwidth. The performance of analog-to-digital converters, data cables and storage devices are main possible bottlenecks to achieve a large sounding bandwidth.

For any sounder, an antenna array is necessary at the Tx/Rx side to capture the spatial channel characteristics. Arrays for sounding can be classified into three categories as illustrated in Fig.\,\ref{fig:thz_sounding}. 
By either mechanically rotating a high-gain antenna, e.g., a horn antenna, to different directions or moving a semi-omnidirectional antenna to different positions, a virtual antenna array can be realized. 
Since only one radio-frequency (RF) chain is required, it is straightforward and cost-effective to apply virtual antenna arrays for THz channel measurements. 
However, the time-consuming mechanical rotations and movements make them only applicable for static THz channels. 
Consequently, virtual antenna arrays are more often combined with VNA-based channel sounding. 
Alternatively, one can develop a real antenna array to perform \addnew{fully} parallel sounding. This means that \addnew{each antenna element has one RF-chain} and all the channels of all the antenna-pairs are measured at the same time. 
This is favorable for dynamic THz channels. However, the number of RF chains, affecting the complexity and cost of the channel sounder, scale with the number of antenna elements. 
\addnew{Moreover, more antenna elements usually mean a smaller bandwidth because the speed of data streaming in the sounder is limited.} 
A good strategy is to use a switched antenna array. 
In this case, a switch matrix is exploited to activate one antenna element at one time, to keep the advantage of one RF-chain without decreasing the bandwidth. 
Moreover, since the electronic switching is very fast, it is often possible to finish one measurement round within a period during which the propagation mechanisms are kept constant. 
Nevertheless, the switch matrix for a large-scale antenna array can introduce significant power loss because of its multi-stage structure. 
Therefore, post-processing is important to harvest beamforming gain of the antenna array to increase the dynamic range. 
Overall, it is crucial to find out the key channel characteristics to be investigated for targeted THz scenarios, so that a proper sounding scheme and array structure can be chosen. Note that extensive measurement data are usually required to achieve sufficient statistics. \addnew{Moreover, it is crucial to gurantee the repeatability and reproducibility. Channel measurements should be conducted along the same route or same positions several times, and the configuration, method, framework, etc. should be clearly explained or publicized so that others can replicate the same experiments.} 

\subsection{Channel parameter estimation}

\begin{figure}
  \centering
  \psfrag{data1}[l][l][0.6]{Measured channel}
  \psfrag{data2}[l][l][0.6]{Estimated path \#1}
  \psfrag{data3}[l][l][0.6]{Estimated path \#2}
  \psfrag{data4}[l][l][0.6]{Estimated path \#3}
  \psfrag{data7}[l][l][0.6]{Residual component}
  \psfrag{Delay}[c][c][0.8]{Delay [s]}
  \psfrag{Power}[c][c][0.8]{Normalized power [dB]}
  \includegraphics[width=0.47\textwidth]{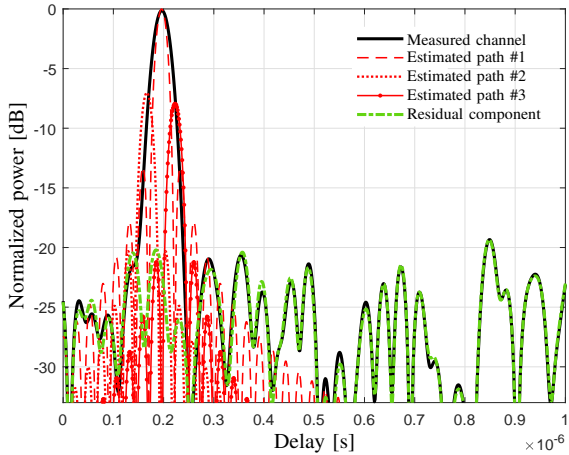}
  \caption{The black curve is a measured CIR of a pure LoS channel with only one propagation path. By (wrongly) assuming that the system response of the channel sounder is frequency-flat, at least 5 paths are extracted so that the residual component (green curve) becomes noisy. The first three estimated paths (red curves) are plotted. The estimated power and delay of path \#1 is consistent with the LoS path as demonstrated by the aligned peaks. However, powers and delays of the rest ghost-paths are all erroneous. \label{fig:sage_paths}}
 \end{figure}

It is essential to accurately estimate THz channel parameters, including delays, angles, etc. of MPCs, from the measurement data. 
High-resolution parameter estimation (HRPE) algorithms based on the maximum likelihood estimation principle can be applied. 
However, apart from additive white Gaussian noise, the system response of a channel sounder are inevitably embedded in the measurement data. 
As the bandwidth of THz sounding signals becomes larger, frequency selectivity of the sounder response becomes non-negligible. 
Moreover, spherical wavefront propagation necessitates the knowledge of 3-D patterns of individual antenna elements since the angles of a same path at different antennas can be different. It is also necessary to estimate the correct powers of MPCs. 
Therefore, it becomes critical to characterize and calibrate the system response to avoid systematic estimation errors for THz channels. 
For example, as illustrated in Fig.\,\ref{fig:sage_paths}, if the frequency-selective system response is not considered, erroneous paths, the so-called ghost paths, can emerge.   


The system response can be divided into two parts when calibrated. The first part is due to the antenna radiation patterns, while the second part is due to all the other components including cables, power amplifiers, filters, converters, etc. of the channel sounder. The latter part is usually characterized by directly connecting the Tx and Rx to obtain the back-to-back channel impulse responses or channel transfer functions. \addnew{It is also necessary to check whether the responses are highly independent on the transmission power.} 
For a THz channel sounder with multiple RF-chains, it is necessary to calibrate all the chain-pairs. 
On the other hand, an antenna array is usually characterized in an anechoic chamber, where the 3-D responses of antenna elements at discrete angle support are measured and stored for the reconstruction of responses of the antenna array at an arbitrary direction. 
One method to do this is to directly perform interpolations based on the measured patterns. However, this is a non-analytic solution hindering the usage of gradient-based optimization in the channel parameter estimation step. An alternative analytic way is to exploit the spherical-harmonics that are orthogonal basis functions defined on a sphere. 
By projecting the measured antenna patterns onto spherical-harmonics, i.e., a forward transform, the coefficients of the spherical-harmonics contained in the measured patterns can be obtained. The array responses at an arbitrary direction can then be recovered using an inverse transform, i.e., a linear combination of spherical-harmonics weighted by their corresponding coefficients. 
Moreover, an extensive set of measured data can be stored compactly, consisting of only principal spherical-harmonic components. 
A third method is the effective aperture distribution function (EADF). Similarly, the EADF transforms the measured antenna patterns via 2-D discrete Fourier transforms to obtain the spectra in the spatial frequency domain. 
The EADF has several advantages. For example, both the forward and inverse discrete Fourier transforms can be efficiently implemented. The principle components can also be utilized to compress the measurement data. In addition, the EADF can be easily extended with frequency domain for a wideband characterization of an antenna array using 3-D discrete Fourier transforms, which well fits the ultrawideband nature of THz channels.

\subsection{Clustering and channel modeling}

\begin{figure}
  \psfrag{Delay [s]}[c][c][0.6]{Delay [s]}
  \psfrag{Azimuth [deg]}[c][c][0.6]{Azimuth [$^\circ$]}
  \psfrag{Power (db)}[c][c][0.6]{Power [dB]}
  \subfigure[]{\includegraphics[width=0.48\textwidth]{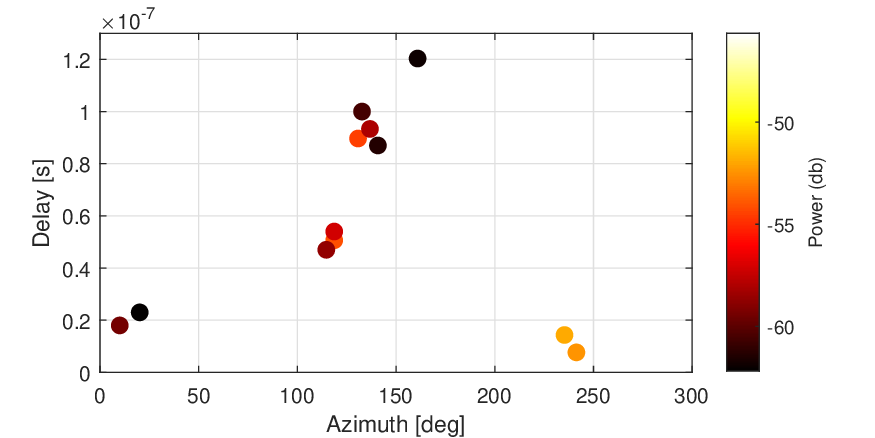}}
  \subfigure[]{\includegraphics[width=0.45\textwidth]{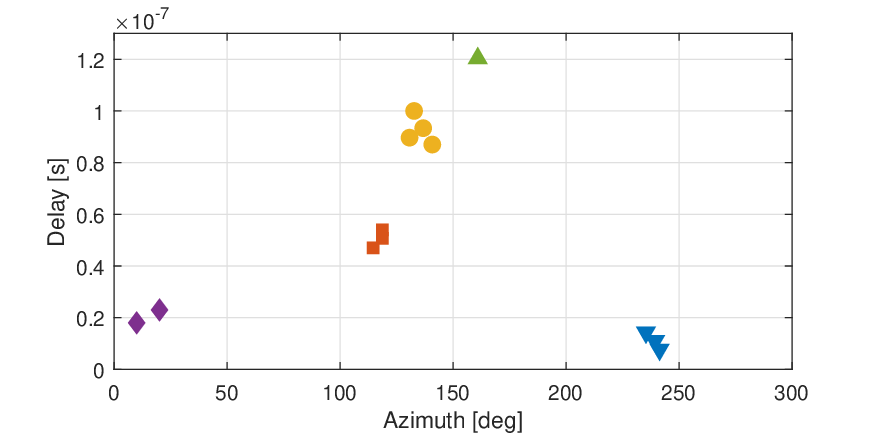}}
  \caption{Grouping estimated MPCs as shown in (a) to several clusters (with different markers) as shown in (b).\label{fig:clustering}}
 \end{figure}

Before establishing channel models, MPCs with similar parameters, e.g., delays and angles, usually need to be grouped as individual clusters as depicted in Fig.\,\ref{fig:clustering}. 
The concept of clusters is essential to balance the model complexity and accuracy. 
There are several clustering algorithms that can be applied for THz channels, typically including K-means, threshold-based and Gaussian-mixture-model algorithms. K-means and threshold-based algorithms rely on the multipath-component-distance, which is calculated by checking the differences of MPCs in the delay and angular domains. The multipath-component-distance is a more feasible distance metric compared to the Euclidean distance since it considers the different properties of different channel parameters. In the K-means method, one needs to first initialize several cluster centroids, assign the nearest MPCs to the corresponding centroids and update the centroids for next iterations until convergence. The artificial parameter of K-means is the number of clusters. Usually, the K-means algorithm is performed for a channel snapshot with different numbers of clusters, and the final clustering result is obtained by checking the compactness of clusters and the separation among clusters. In the threshold-based method, MPCs that are within a distance-threshold to the currently strongest MPC are grouped as the same cluster and removed to continue the grouping of the next cluster. The only artificial parameter herein is the threshold. Similarly, one needs to find the optimal threshold via trials. K-means and threshold-based algorithms do not assume any prior distributions of the MPCs. Whereas in the Gaussian-mixture-model algorithm, MPCs inside one cluster are assumed to have a Gaussian distribution, based on which the MPCs are grouped according to the maximum-likelihood-estimation principle. It is hard to claim which clustering algorithm that is the best, which depends on the specific channel realization. In addition, to study dynamic THz channels, clusters also need to be tracked and associated along space, time and the array aperture. Based on the clustering and tracking results, THz channel characteristics such as path loss, shadowing, small scale fading, intra-cluster characteristics, inter-cluster characteristics, birth-death behavior of clusters, etc., as summarized in Table\,\ref{tab:summarization}, can be modelled statistically with distributions. Non-experts in channel modeling can take these distributions to easily simulate the target THz channels for their own investigations.

\begin{table*}[h]
  \centering
    \caption{\addnew{A brief summary of existing THz channel measurements and future research directions.}}
    \scalebox{1}{\textcolor{black}{\footnotesize
    \begin{tabular}{lp{4cm}p{4cm}p{4cm}p{4cm}}
    \hline
   {} & Existing works & Missing works & Future directions  \\ \hline
   Channel sounding &  Mostly VNA-based, static scenarios, short distances & Dynamic double-directional dual-polarized channel sounding  &  RIS-assisted, switched array based, or fast beam-scan based channel sounders \\ \hline
                        Channel parameter estimation & Mainly based on power spectra & HRPE algorithms &  Low-complexity HRPE algorithms \\ \hline
                       Channel properties/models  & Large- and small-scale fading, delay and angle spreads, clusters, blockage  &  Spatial consistency  &  Physical scatterer based channel models, AI-assisted channel modeling     \\  \hline 
    \end{tabular}}}
    \label{tab:summarizationb}
    \end{table*}

\section{Existing THz channel measurements\label{sec:existingwork}}

In general, THz channel measurements are still scarce in the literature.  
Channel sounders utilized in most of the existing works were based on VNAs and up/down-converters. 
Measurement distances were usually not large, e.g., up to 10\,m, due to cable connections and high power losses in the cables. Nevertheless, using the technique of radio-over-fiber, measurement distance can be extended effectively, e.g., to be more than 100\,m as shown in \cite{9790802}. 
Time-domain channel sounders were also exploited in some works, where the measurement distance can be naturally larger, e.g., up to 100\,m \cite{9792355}.  
With single-input single-output configuration, channel characteristics investigated usually include path loss, delay spread and power delay profile. 
To capture spatial THz channel characteristics, almost all the relevant works used the scheme of mechanically rotating high-gain horn antennas. 
MPCs, delay spreads, angular spreads, clusters, etc. can be further investigated. However, due to that the beamwidths of high-gain THz antennas are narrow and many rotation steps are needed, it is difficult to measure dynamic THz channels with Tx, Rx or scatterers moving even at low speeds. 
Consequently, almost all the existing THz channel measurements were done in a static manner. For example, even for vehicle-to-vehicle THz channels, the measurements were actually conducted inside a room with static cars \cite{9403881}. 
\addnew{The vehicle-body blockage is found to be highly dependent on the antenne heights. It can be up to 50\,dB with antennas at the engine level. As the heights increase to the windshield level, the blockage loss can drop by almost 20\,dB and then grow again. When the antenna heights further increase, reflection on the rooftop can result in negative additional-loss.}
In addition, there are some works studying diffraction, penetration, scattering and reflection in the THz band. Theoretical models, e.g., the directive scattering model for scattering \cite{8761205}, were analyzed with measurement data. Accurate understanding of these fundamental mechanisms are important for both simulation tools and the interpretations of measured observations.  

Measurements show that the path loss exponent at THz band is close to 2 in most cases, especially in LoS scenarios. The path loss intercept, i.e., path loss at 1\,m, increases rapidly with increasing frequencies. Blockage loss due to walls, cars, etc. can be large on the order of several tens of dB, which means that severe shadow fading may happen. 
Moreover, since the THz wavelengths are very short, a small movement can change constructive additions of MPCs to destructive additions, resulting in deep fades. 
Scattering depends on surfaces and impinging angles and can be well characterized with the directive scattering model \cite{9013236}. The scattered power relative to the reflected power increases with frequency and surface roughness, and the reflection coefficient linearly increases as the incident angle increases. Smooth surfaces like dry walls may be modelled as simple reflective surfaces, especially when the incident angle is large. As for diffraction, measurements show that it can be well modelled with the uniform geometrical theory of diffraction.
Due to the severe path loss and blockage loss, the number of MPCs and clusters are usually not large. Cluster spreads are also observed to be small. However, it is possible that the composite angular spread is large due to local scatterers at different angles \cite{9790802}. Although evidences show that the THz channels are likely sparse, it is probably due to the low dynamic ranges of the channel sounders. With a large scale antenna array. i.e., a large effective area, applied in the channel sounding, many more paths may be revealed. These paths can be exploited in THz communications with UM-MIMO. The sparsity of THz channels should not be affirmed too early.


\section{Future research directions\label{sec:futuredic}}


\textit{Novel channel sounders}: 
Channel sounders with longer measurement distances and larger dynamic ranges are required for mid-to-long distance THz applications. 
Possible solutions can be exploiting advanced THz devices, e.g., traveling wave tubes as power amplifiers providing an output power up to 40\,dBm \cite{Paoloni2020}, and designing proper signals with high sequence gains for time-domain channel sounders.  
Moreover, switched antenna arrays or real-time beam-rotation, e.g., adjusting the beam of a horn antenna with a reconfigurable intelligent surface (RIS), are essential to establish dynamic THz channel models with spatial consistency. 
Further, no large-scale arrays have been applied in measurements so far, which hinders the understanding of several important THz channel characteristics including spherical-wave propagation, channel evolution across the array and the correspondence between channels and physical scatterers. 
From this perspective, a large switched antenna array with time-domain channel sounding would be the most preferred design of a THz channel sounder for joint sensing and communication applications. In addition, little is known about electromagnetic parameters of various materials at THz band, which hinders the application of simulation-based channel modeling.  Measurements tackling this problem are needed.

\textit{Low-complexity HRPE algorithms}:  The ultra-wide bandwidths and large-scale antenna arrays can bring significant advantages to distinguish MPCs. However, the complexities of the HRPE algorithms can increase fatally due to many more parameter domains and larger sizes of measurement data. For example, for a dynamic double-directional ultra-wideband UM-MIMO THz channel measurement, channel parameters to be estimated for one MPC include delay, Doppler frequency, azimuth/elevation of departure/departure, azimuth/elevation of arrival/departure, spherical distance of arrival/departure and polarimetric amplitudes, i.e., 16 parameters in total. Moreover, the narrowband and plane-wave assumptions are not valid anymore, which makes it difficult to decouple parameters to decompose the problem into several one-dimensional estimation problems. Therefore, novel HRPE algorithms with low complexities are indispensable. One promising approach is exploiting delay trajectories of MPCs across an UM-MIMO array to achieve fast initializations and effective interference cancellations, so that the searching space can be reduced significantly in the channel parameter estimation, e.g., as done in \cite{8713575}. 


\textit{Spatially consistent THz channel models}: THz channel models applicable for communication, sensing and localization in 6G should well describe the geometry to address the correspondence between the physical environment and the THz channels, describe the spatial consistency to present a smooth channel change without discontinuities introduced by small change in Tx and Rx, and describe the birth-death behavior of MPCs, clusters and physical scatterers for tracking and identifying objects in the environment. \addnew{Moreover, continuos evolution of channels across large arrays, such as UM-MIMO and RIS, is also crucial.} However, existing THz channel models are far from meeting the requirements. Besides lacking THz channel sounders to measure these characteristics, a new THz channel modeling framework is also required. The most promising one, simply put, is to model the evolution behavior of physical scatterers and add statistical properties to the MPCs generated by the scatterers. 


\textit{AI-assisted THz channel investigation}: Artificial intelligence (AI), especially machine learning techniques, can be applied to many aspects such as propagation scenario identification and channel prediction with deep neural networks, and channel reproduction using generative adversarial networks \cite{9713743}. Besides those, we envision a new promising AI-assisted investigation for THz channels. That is, the spatially THz channels can be inferred from a photo or an optical scan of the environment since they inherently contain the information of physical scatterers in the environment. The challenge is to train the AI-models to map the photo or the scan to the realistic THz channels, which requires not only advanced THz channel sounders but also photo systems or optical systems to obtain massive amount of appropriate measurement data. 

\addnew{Table\,\ref{tab:summarizationb} provides a brief summary of Sects.\,\ref{sec:existingwork} and \ref{sec:futuredic}.}

\section{Conclusions\label{sect:conclusions}}

The new THz band is a key component for sixth-generation (6G) communications. 
Many advanced applications, e.g., virtual reality with short-range, high-throughput and low latency data streaming, can be realized, due to the ultrawide bandwidth available at the THz band. 
Comprehensively understanding the THz propagation channels is an essential and first step to develop the corresponding THz wireless systems. 
6G-compliant THz channel models should be dynamic, spatially consistent, and reveal the correspondence between the physical environment and the channel characteristics. 
In general, THz channel measurements are still scarce in the literature mainly because of lacking proper THz channel sounders. 
Most of the investigations focus on short-range static THz channels. 
New THz channel sounders are critically needed to gain more channel knowledge for different types of THz communications, especially joint communication and sensing in mobile scenarios. Moreover, novel low-complexity THz channel parameter estimation algorithms and physical-scatterer based THz modeling methodologies are required. Artificial intelligence should also be embraced. 


\bstctlcite{IEEEexample:BSTcontrol}
\setlength{\itemsep}{10em}
\renewcommand{\baselinestretch}{0.95}
\bibliographystyle{IEEEtran}
\bibliography{reference}

\begin{thebibliography}{10}
\providecommand{\url}[1]{#1}
\csname url@samestyle\endcsname
\providecommand{\newblock}{\relax}
\providecommand{\bibinfo}[2]{#2}
\providecommand{\BIBentrySTDinterwordspacing}{\spaceskip=0pt\relax}
\providecommand{\BIBentryALTinterwordstretchfactor}{4}
\providecommand{\BIBentryALTinterwordspacing}{\spaceskip=\fontdimen2\font plus
\BIBentryALTinterwordstretchfactor\fontdimen3\font minus
  \fontdimen4\font\relax}
\providecommand{\BIBforeignlanguage}[2]{{%
\expandafter\ifx\csname l@#1\endcsname\relax
\typeout{** WARNING: IEEEtran.bst: No hyphenation pattern has been}%
\typeout{** loaded for the language `#1'. Using the pattern for}%
\typeout{** the default language instead.}%
\else
\language=\csname l@#1\endcsname
\fi
#2}}
\providecommand{\BIBdecl}{\relax}
\BIBdecl

\bibitem{zhang2020channel}
J.-h. Zhang, P.~Tang, L.~Yu, T.~Jiang, and L.~Tian, ``Channel measurements and
  models for {6G}: current status and future outlook,'' \emph{Frontiers of
  information technology \& electronic engineering}, vol.~21, no.~1, pp.
  39--61, 2020.

\bibitem{8674541}
{K. Guan \textit{et al.}}, ``Channel characterization for intra-wagon
  communication at 60 and 300 {GHz} bands,'' \emph{IEEE Transactions on
  Vehicular Technology}, vol.~68, no.~6, pp. 5193--5207, 2019.

\bibitem{rajatheva2020scoring}
{N. Rajatheva \textit{et al.}}, ``Scoring the terabit/s goal: Broadband
  connectivity in {6G},'' \emph{arXiv:2008.07220}, 2020.

\bibitem{9216613}
A.~Faisal, H.~Sarieddeen, H.~Dahrouj, T.~Y. Al-Naffouri, and M.-S. Alouini,
  ``Ultramassive {MIMO} systems at terahertz bands: Prospects and challenges,''
  \emph{IEEE Vehicular Technology Magazine}, vol.~15, no.~4, pp. 33--42, 2020.

\bibitem{9047145}
{B. Peng \textit{et al.}}, ``Channel modeling and system concepts for future
  terahertz communications: Getting ready for advances beyond {5G},''
  \emph{IEEE Vehicular Technology Magazine}, vol.~15, no.~2, pp. 136--143,
  2020.

\bibitem{9482537}
{O. Li \textit{et al.}}, ``Integrated sensing and communication in {6G} a
  prototype of high resolution {THz} sensing on portable device,'' in
  \emph{Joint European Conference on Networks and Communications and 6G
  Summit}, 2021, pp. 544--549.

\bibitem{9790802}
J.~Gomez-Ponce, N.~A. Abbasi, A.~E. Willner, C.~J. Zhang, and A.~F. Molisch,
  ``Directionally resolved measurement and modeling of {THz} band propagation
  channels,'' \emph{IEEE Open Journal of Antennas and Propagation}, vol.~3, pp.
  663--686, 2022.

\bibitem{8255203}
S.~Rey, J.~M. Eckhardt, B.~Peng, K.~Guan, and T.~Kürner, ``Channel sounding
  techniques for applications in {THz} communications: A first correlation
  based channel sounder for ultra-wideband dynamic channel measurements at 300
  {GHz},'' in \emph{9th International Congress on Ultra Modern
  Telecommunications and Control Systems and Workshops (ICUMT)}, 2017, pp.
  449--453.

\bibitem{9792355}
K.-W. Kim, M.-D. Kim, J.-J. Park, J.~Lee, and H.-K. Kwon, ``Path loss and
  multipath profiles in a street canyon based on 253 {GHz} measurements,'' in
  \emph{14th Global Symposium on Millimeter-Waves \& Terahertz (GSMM)}, 2022,
  pp. 113--115.

\bibitem{9403881}
J.~M. Eckhardt, V.~Petrov, D.~Moltchanov, Y.~Koucheryavy, and T.~Kürner,
  ``Channel measurements and modeling for low-terahertz band vehicular
  communications,'' \emph{IEEE Journal on Selected Areas in Communications},
  vol.~39, no.~6, pp. 1590--1603, 2021.

\bibitem{8761205}
S.~Ju, S.~H.~A. Shah, M.~A. Javed, J.~Li, G.~Palteru \emph{et~al.},
  ``Scattering mechanisms and modeling for terahertz wireless communications,''
  in \emph{IEEE International Conference on Communications (ICC)}, 2019, pp.
  1--7.

\bibitem{9013236}
Y.~Xing, O.~Kanhere, S.~Ju, and T.~S. Rappaport, ``Indoor wireless channel
  properties at millimeter wave and sub-terahertz frequencies,'' in \emph{IEEE
  Global Communications Conference (GLOBECOM)}, 2019, pp. 1--6.

\bibitem{Paoloni2020}
{C. Paoloni \textit{et al.}}, ``Toward the first {D-band} point to multipoint
  wireless system field test,'' in \emph{Joint European Conference on Networks
  and Communications \& 6G Summit}, 2021, pp. 55--59.

\bibitem{8713575}
X.~Cai and W.~Fan, ``A complexity-efficient high resolution propagation
  parameter estimation algorithm for ultra-wideband large-scale uniform
  circular array,'' \emph{IEEE Transactions on Communications}, vol.~67, no.~8,
  pp. 5862--5874, 2019.

\bibitem{9713743}
{C. Huang \textit{et al.}}, ``Artificial intelligence enabled radio propagation
  for communications-part {II}: Scenario identification and channel modeling,''
  \emph{IEEE Transactions on Antennas and Propagation}, vol.~70, no.~6, pp.
  3955--3969, 2022.

\end{thebibliography}

\vspace{0cm}

\begin{IEEEbiography}[{\includegraphics[width=1in,height=1.25in,clip,keepaspectratio]{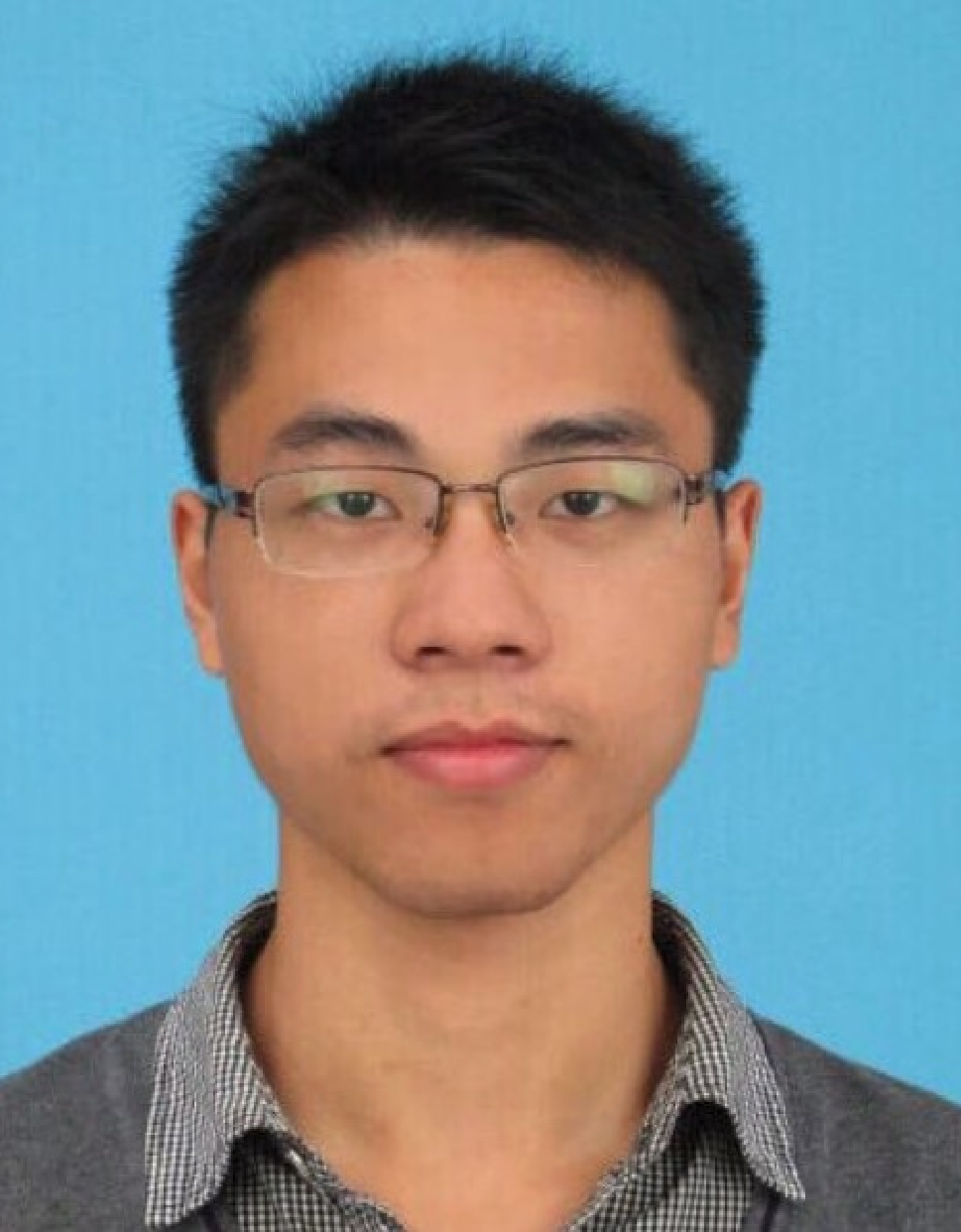}}]{Xuesong Cai} received the B.S. degree and the Ph.D. degree (Hons.) from Tongji University, Shanghai, China, in 2013 and 2018, respectively. In 2015, he conducted a three-month internship with Huawei Technologies, Shanghai, China. He was also a Visiting Scholar at Universidad Politècnica de Madrid, Madrid, Spain in 2016. From 2018-2022, he conducted several postdoctoral stays at Aalborg University, Aalborg, Denmark, Nokia Bell Labs, Aalborg, Denmark, and Lund University, Lund, Sweden. 

Since December 2022, he has been an Assistant Professor of wireless communication in the Department of Electrical and Information Technology, Lund University. His research interests include radio propagation, high-resolution parameter estimation, over-the-air testing, resource optimization, and radio-based localization for 5G and beyond wireless systems.

Dr. Cai was a recipient of the China National Scholarship (the highest honor for Ph.D. Candidates) and the Excellent Student in 2016, the Excellent Student and the ``ZTE Fantastic Algorithm'' award in 2017, the Outstanding Doctorate Graduate awarded by the Shanghai Municipal Education Commission in 2018, the Marie Skłodowska-Curie Actions (MSCA) ``Seal of Excellence'' in 2019, the MSCA Postdoctoral Fellowship (ranking top 1.2\%, success rate 14\%) founded by EU and the Starting Grant (success rate 12\%) funded by the Swedish Research Council in 2022. He was also selected by the ``ZTE Blue Sword-Future Leaders Plan'' in 2018 and the ``Huawei Genius Youth Program'' in 2021. 
\end{IEEEbiography}

\begin{IEEEbiography}[{\includegraphics[width=1in,height=1.25in,clip,keepaspectratio]{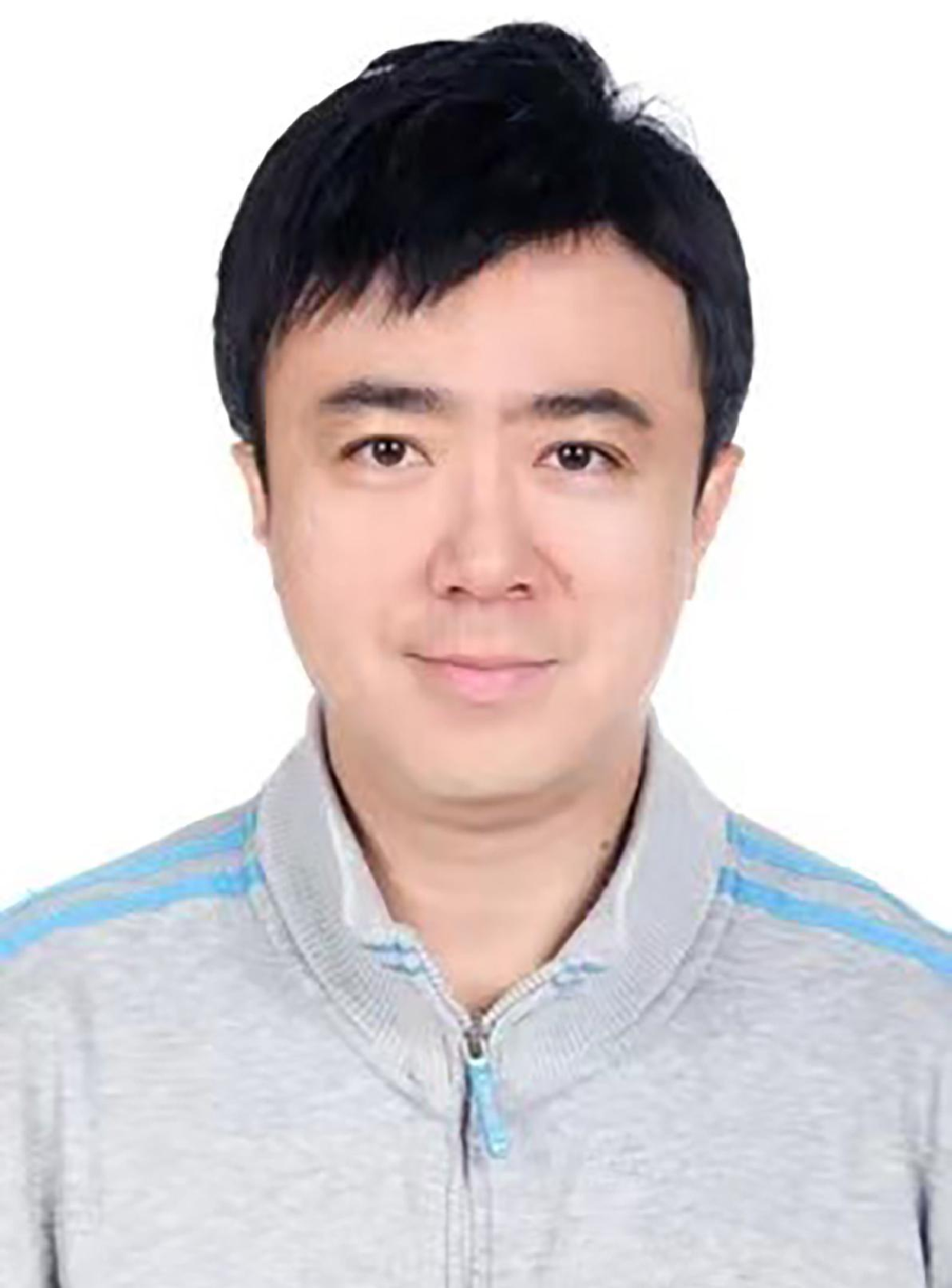}}]{Xiang Cheng} (S’05-M’10-SM’13-F’22) received the Ph.D. degree jointly from Heriot-Watt University and the University of Edinburgh, Edinburgh, U.K., in 2009. He is currently a Boya Distinguished Professor of Peking University. His general research interests are in areas of channel modeling, wireless communications, and data analytics, subject on which he has published more than 280 journal and conference papers, 9 books, and holds 17 patents. Prof. Cheng is a Distinguished Young Investigator of China Frontiers of Engineering, a recipient of the IEEE Asia Pacific Outstanding Young Researcher Award in 2015, a Distinguished Lecturer of IEEE Vehicular Technology Society, and a Highly Cited Chinese Researcher in 2020. He was a co-recipient of the 2016 IEEE JSAC Best Paper Award: Leonard G. Abraham Prize, and IET Communications Best Paper Award: Premium Award. He has also received the Best Paper Awards at IEEE ITST’12, ICCC’13, ITSC’14, ICC’16, ICNC’17, GLOBECOM’18, ICCS’18, and ICC’19. He has served as the symposium lead chair, co-chair, and member of the Technical Program Committee for several international conferences. He is currently a Subject Editor of IET Communications and an Associate Editor of the IEEE Transactions on Wireless Communications, IEEE Transactions on Intelligent Transportation Systems, IEEE Wireless Communications Letters, and the Journal of Communications and Information Networks. In 2021, he was selected into two world scientist lists, including World’s Top 2\% Scientists released by Stanford University and Top Computer Science Scientists released by Guide2Research.
\end{IEEEbiography}

\begin{IEEEbiography}[{\includegraphics[width=1in,height=1.25in,clip,keepaspectratio]{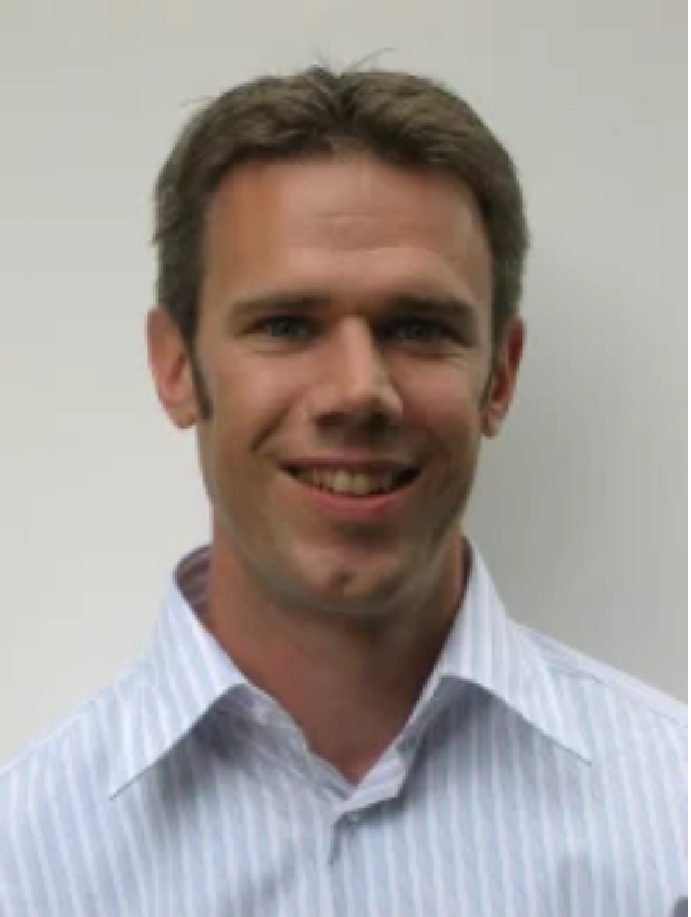}}]{Fredrik Tufvesson} (Fellow, IEEE) 
received the Ph.D. degree from Lund University, Lund, Sweden, in 2000.

After two years at a startup company, he joined the Department of Electrical and Information Technology, Lund University, where he is currently a Professor of radio systems. He has authored around 100 journal articles and 150 conference papers. His main research interest is the interplay between the radio channel and the rest of the communication system with various applications in 5G/B5G systems, such as massive multiple-input multiple-output (MIMO), mmWave communication, vehicular communication, and radio-based positioning.

Dr. Tufvesson’s research has been awarded the Neal Shepherd Memorial Award for the Best Propagation Paper in the IEEE
TRANSACTIONS ON VEHICULAR TECHNOLOGY and the IEEE Communications Society Best Tutorial Paper Award.

\end{IEEEbiography}




  



\end{document}